\documentclass{webofc}
\usepackage[varg]{txfonts}   % Web of Conferences font
\usepackage{epsf,epsfig,float,latexsym,rotating,graphicx,slashed,bm,ulem}
\usepackage{amsmath,amssymb,amsfonts}
\usepackage[dvipsnames,usenames]{color}
\usepackage{array,multirow,booktabs}
\usepackage[utf8]{inputenc}
\usepackage[sort&compress]{natbib}
\newcommand{\beq}{\begin{equation}}
\newcommand{\eeq}[1]{\label{#1}\end{equation}}
\newcommand{\beqa}{\begin{eqnarray}}
\newcommand{\eeqa}[1]{\label{#1}\end{eqnarray}}
\newcommand{\eeqan}{\end{eqnarray}}

\newcommand{\CSnlo}{$\rm P_{\rm NLO}$}

\newcommand{\GOi}{$\rm M_{\rm I}$}
\newcommand{\GOii}{$\rm M_{\rm II}$}

\newcommand{\IHWnlo}{$\rm KM_{\rm NLO}$}
\newcommand{\MMii}{$\rm B_{2}$}
\newcommand{\MMiv}{$\rm B_{4}$}
\makeatletter
\def\gsim{\compoundrel>\over\sim}

\def\compoundrel#1\over#2{\mathpalette\compoundreL{{#1}\over{#2}}}
\def\compoundreL#1#2{\compoundREL#1#2}
\def\compoundREL#1#2\over#3{\mathrel
      {\vcenter{\hbox{$\m@th\buildrel{#1#2}\over{#1#3}$}}}}
\makeatother
\begin{document}
\title{Dynamically generated hadronic states in the $\bar{K}N$ and $\eta N$ coupled-channels interactions}

\author{\firstname{Ale\v{s}} \lastname{Ciepl\'{y}}\inst{1}\fnsep\thanks{\email{cieply@ujf.cas.cz}} 
}

\institute{Nuclear Physics Institute, 250 68 \v{R}e\v{z}, Czechia 
}

\abstract{%
We compare and review the theoretical predictions for the $\bar{K}N$ and $\eta N$ elastic amplitudes 
where sizable variations are found among the considered approaches, especially at subthreshold 
energies relevant for studies of kaonic atoms and meson-nuclear quasi-bound states. 
Conditions for an appearance of dynamically generated states in meson-baryon multi-channel 
interactions are established and discussed for the $\bar{K}N$ and $\eta N$ systems.
}
\maketitle
%
%%%%%%%%%%%%%%%%%%%%%%%%%%%%%%%%%%%%%%%%%%%%%%%%%%%%%%%%%%%%%%%%%%%%%%%%%%%%%%%%%%%%%%%%%%%%%%%%%%%%%
\section{Introduction}
\label{sec-intro}
%%%%%%%%%%%%%%%%%%%%%%%%%%%%%%%%%%%%%%%%%%%%%%%%%%%%%%%%%%%%%%%%%%%%%%%%%%%%%%%%%%%%%%%%%%%%%%%%%%%%%
%
A modern treatment of low-energy meson-nucleon interactions is based on effective chiral Lagrangians 
that implement the QCD dynamics and symmetries in a regime where the large running coupling constant 
makes the perturbative approach intractable. A good number of chiral approaches have been developed 
to describe the s-wave interactions of the pseudoscalar meson octet ($\pi$, $K$, $\eta$) with the ground 
state baryon octet ($N$, $\Lambda$, $\Sigma$, $\Xi$), see e.g. \cite{Kaiser:1995eg, Oset:1997it, 
Jido:2003cb, Ikeda:2012au, Cieply:2011nq, Guo:2012vv, Mai:2014xna} for papers dealing 
with the $\bar{K}N$ system and \cite{Kaiser:1995cy, Inoue:2001ip, Mai:2012wy, Cieply:2013sya} 
for those related to the $\eta N$ one. Unfortunately, the presence of resonances hinders the use 
of perturbative chiral expansion. This problem is standardly overcome by summing the presumably 
dominant part of the perturbation series by means of either the Lippmann-Schwinger or Bethe-Salpeter 
coupled-channels equations. The effective potentials that enter these equations are then constructed 
to match the meson-baryon scattering amplitudes derived from the underlying chiral Lagrangian 
up to a given chiral order. Free parameters of the models are fixed partly by relating them 
to meson and baryon properties, and partly by fitting the available experimental data. 

In our contribution we reflect briefly on the current status of theoretical approaches 
to the $\bar{K}N$ and $\eta N$ systems and discuss the conditions for an appearance of 
dynamically generated resonances in such strongly interacting multi-channel systems. 
Both systems are characterized by well established resonant states observed very close 
to the channel thresholds. In the strangeness $S=-1$ sector it is the $\Lambda(1405)$ resonance 
lying about 30 MeV below the $\bar{K}N$ threshold while the low energy $\eta N$ interaction 
representing the $S=0$ sector is dominated by the $N^{*}(1535)$ resonance found about 50 MeV 
above the elastic channel threshold. The resonances have a big impact on dynamics of the $\bar{K}N$ 
and $\eta N$ systems not only in the free space but they also influence significantly 
the propagation of the $\bar{K}$ and $\eta$ mesons in nuclear matter at energies close to threshold. 
In particular, the findings presented in our contribution are also relevant for studies of 
kaonic atoms and for a possible formation of $\bar{K}$-nuclear and $\eta$-nuclear quasi-bound states
\cite{Cieply:2011fy, Cieply:2013sga, Friedman:2016rfd, Hrtankova:2017zxw}.

In the next section, we demonstrate the ability of several up-to-date chiral approaches to reproduce 
selected experimental data and show the ambiguities of the theoretical models when making 
predictions in areas not so well restricted by the data. We then turn our attention to looking 
at the origin of the dynamically generated resonances relating them to the poles of the scattering 
amplitudes and discuss conditions of their existence. The paper is closed with a brief summary.

%
%%%%%%%%%%%%%%%%%%%%%%%%%%%%%%%%%%%%%%%%%%%%%%%%%%%%%%%%%%%%%%%%%%%%%%%%%%%%%%%%%%%%%%%%%%%%%%%%%%%%%
\section{Model predictions and comparisons}
\label{sec-models}
%%%%%%%%%%%%%%%%%%%%%%%%%%%%%%%%%%%%%%%%%%%%%%%%%%%%%%%%%%%%%%%%%%%%%%%%%%%%%%%%%%%%%%%%%%%%%%%%%%%%%
%
%
%%%%%%%%%%%%%%%%%%%%%%%%%%%%%%%%%%%%%%%%%%%%%%%%%%%%%%%%%%%%%%%%%%%%%%%%%%%%%%%%%%%%%%%%%%%%%%%%%%%%%
\subsection{$\bar{K}N$ amplitudes}
\label{sec:KbarN}
%%%%%%%%%%%%%%%%%%%%%%%%%%%%%%%%%%%%%%%%%%%%%%%%%%%%%%%%%%%%%%%%%%%%%%%%%%%%%%%%%%%%%%%%%%%%%%%%%%%%%
%
The meson-baryon channels involved in the $S=-1$ sector 
($\pi\Lambda$, $\pi\Sigma$, $\bar{K}N$, $\eta\Lambda$, $\eta\Sigma$ and $K\Xi$) 
have the channel thresholds spread over a large interval of energies from $1250$ to $1810$ MeV. 
However, since we aim at description of $\bar{K}N$ interactions for energies at and below 
the elastic threshold the relevant experimental data that determine the fitted model parameters
are the $K^{-}p$ threshold branching ratios, low energy elastic and inelastic $K^{-}p$ cross-sections 
(see e.g. \cite{Kaiser:1995eg} for references)
and 1s level characteristics (the shift and width due to strong interaction) of kaonic hydrogen, 
measured recently with quite high accuracy \cite{Bazzi:2011zj}. This set of experimental data is reproduced 
equally well by all current theoretical approaches that include the NLO corrections 
to the leading order in the chiral expansion. An overview of the models and their direct comparison 
was discussed in \cite{Cieply:2016jby}. Here we follow the notation introduced there and label 
the models by locations of their authors referring to the Kyoto-Munich (\IHWnlo{}) \cite{Ikeda:2012au}, 
Prague (\CSnlo{}) \cite{Cieply:2011nq}, Murcia (\GOi{} and \GOii{}) \cite{Guo:2012vv} 
and Bonn (\MMii{} and \MMiv{}) \cite{Mai:2014xna} approaches.

\begin{figure}[ht]
\centering
\resizebox{0.9\textwidth}{!}
{\includegraphics{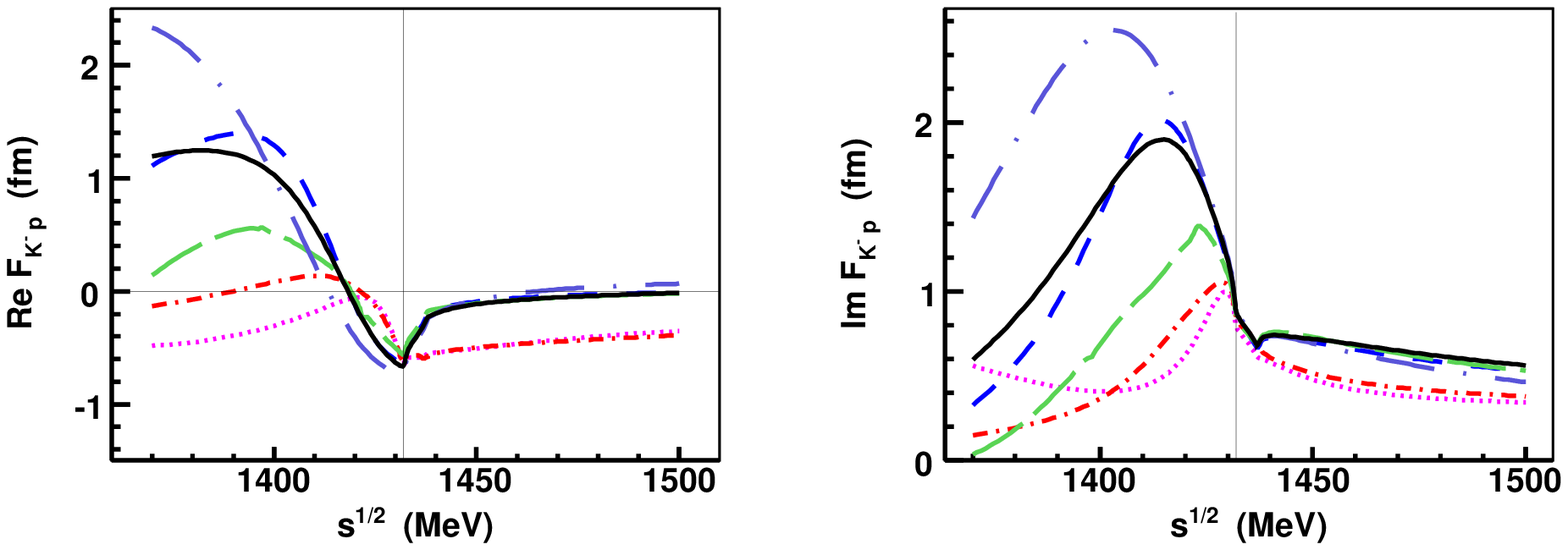}
}\\
\resizebox{0.9\textwidth}{!}
{\includegraphics{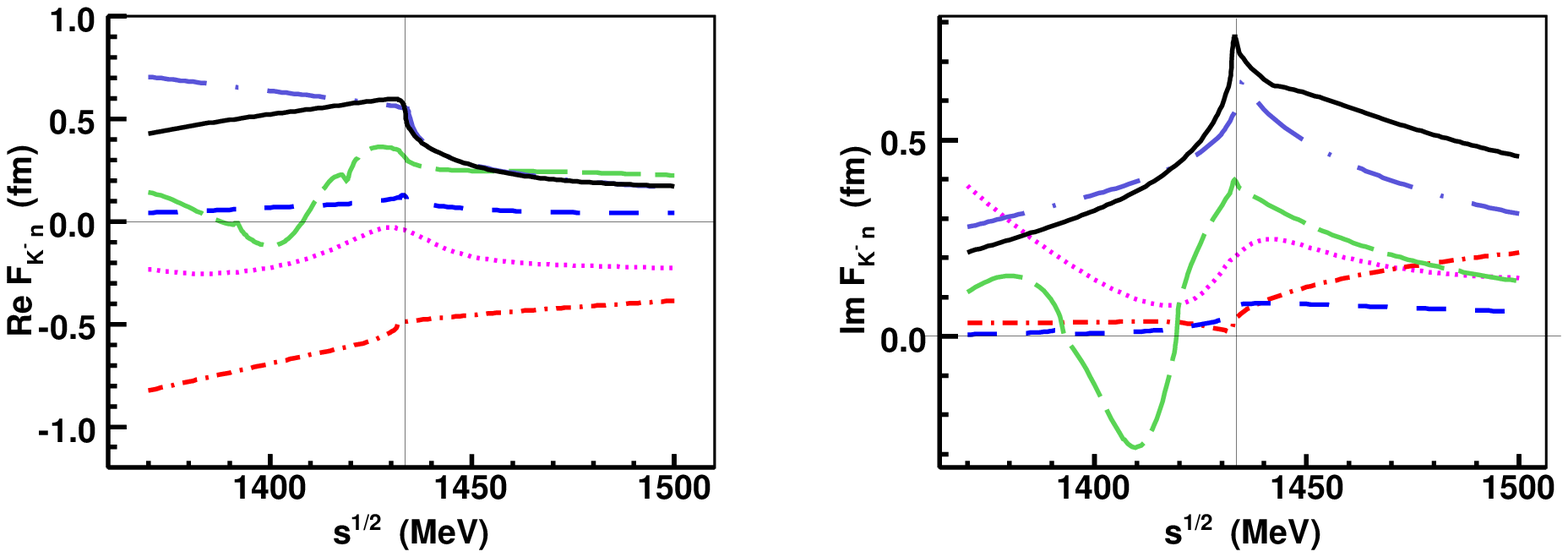}
}
\caption{The $K^{-}p$ (top panels) and $K^{-}n$ (bottom panels) elastic scattering amplitudes 
generated by the NLO approaches considered in our work. The various lines refer to the models:  
\MMii{} (dotted, purple), \MMiv{} (dot-dashed, red), \GOi{} (dashed, blue), 
\GOii{} (long-dashed, green), \CSnlo{} (dot-long-dashed, violet), \IHWnlo{} (continuous, black).
The thin vertical lines mark the $\bar{K}N$ threshold.
}
\label{fig:KNampl}       
\end{figure}

In Fig.~\ref{fig:KNampl} taken from \cite{Cieply:2016jby} we demonstrate that the considered 
approaches lead to very different predictions for the $K^{-}p$ amplitude extrapolated 
to sub-threshold energies as well as for the $K^{-}n$ amplitude. The theoretical ambiguities 
observed below the $\bar{K}N$ threshold are much larger then those standardly indicated 
by uncertainty bounds derived from variations of the $K^{-}p$ scattering length
within constraints enforced by the kaonic hydrogen data, see e.g.~Ref.~\cite{Ikeda:2012au}. 
Although the theoretical predictions differ significantly in areas not constrained 
by experimental data the magnitude of these variations has not been anticipated earlier. 
We also note that the $K^{-}p$ amplitude generated by the Bonn models \MMii{} and \MMiv{} 
deviates from the other model predictions even at energies above the threshold due to 
a conceptual difference in a treatment of the s-wave projection. Other differences 
are caused by different fitting procedures and by inclusion of additional experimental data, 
the $\pi\Sigma$ mass distributions in the CLAS photo-production measurement \cite{Moriya:2013eb} 
in case of the Bonn approach, and more data for processes at higher energies around 
the $\Lambda(1670)$ resonance in case of the Murcia approach. 

%%%%%%%%%%%%%%%%%%%%%%%%%%%%%%%%%%%%%%%%%%%%%%%%%%%%%%%
%%%%%%%%%%%%%%%%%%%%%%%%%%%%%%%%%%%%%%%%%%%%%%%%%%%%%%%

%
%%%%%%%%%%%%%%%%%%%%%%%%%%%%%%%%%%%%%%%%%%%%%%%%%%%%%%%%%%%%%%%%%%%%%%%%%%%%%%%%%%%%%%%%%%%%%%%%%%%%%
\subsection{$\eta N$ amplitudes}
\label{sec:etaN}
%%%%%%%%%%%%%%%%%%%%%%%%%%%%%%%%%%%%%%%%%%%%%%%%%%%%%%%%%%%%%%%%%%%%%%%%%%%%%%%%%%%%%%%%%%%%%%%%%%%%%
%
The $\eta N$ interactions that represent the $S=0$ sector were studied in \cite{Cieply:2013sya}. 
In close analogy with the $\bar{K}N$ interactions only the two-body meson-baryon channels involving 
particles from the meson and baryon octets were considered, i.e. the $\pi N$, $\eta N$, $K\Lambda$ 
and $K\Sigma$ channels. The model parameters were fitted to the low energy $S_{11}$ and $S_{31}$ 
partial waves amplitudes and to the $\pi^{-} p \rightarrow \eta n$ reaction total cross section data, 
both taken from the SAID database \cite{SAID}. The quality of the fits was quite good ($\chi^{2}/dof \approx 1.5$) 
and in particular, the structure of the $N^{\star}(1535)$ resonance observed in the $\eta n$ 
production cross sections was nicely reproduced. 

\begin{figure}[ht]
\centering
\resizebox{0.45\textwidth}{!}
{\includegraphics{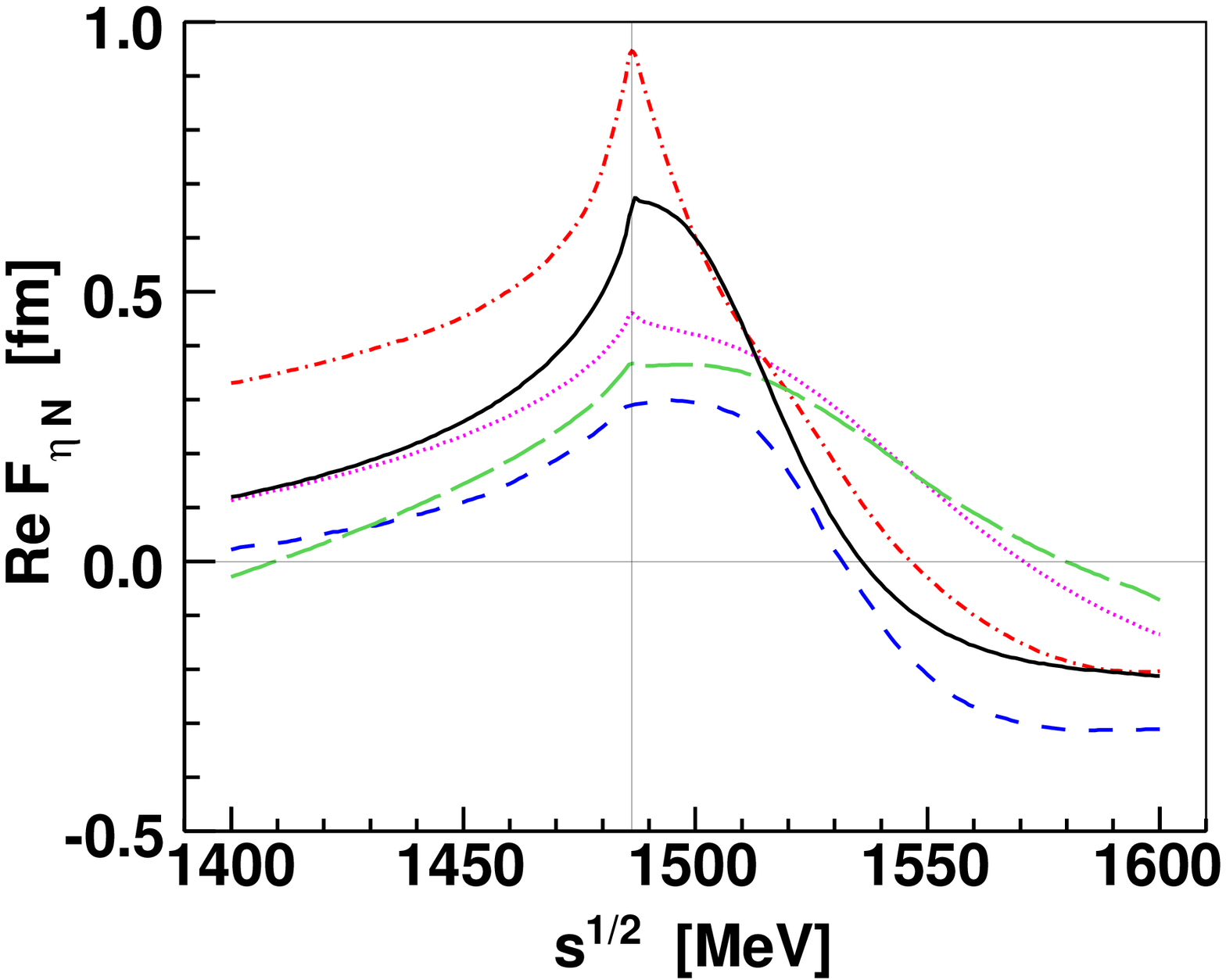}
}
\resizebox{0.45\textwidth}{!}
{\includegraphics{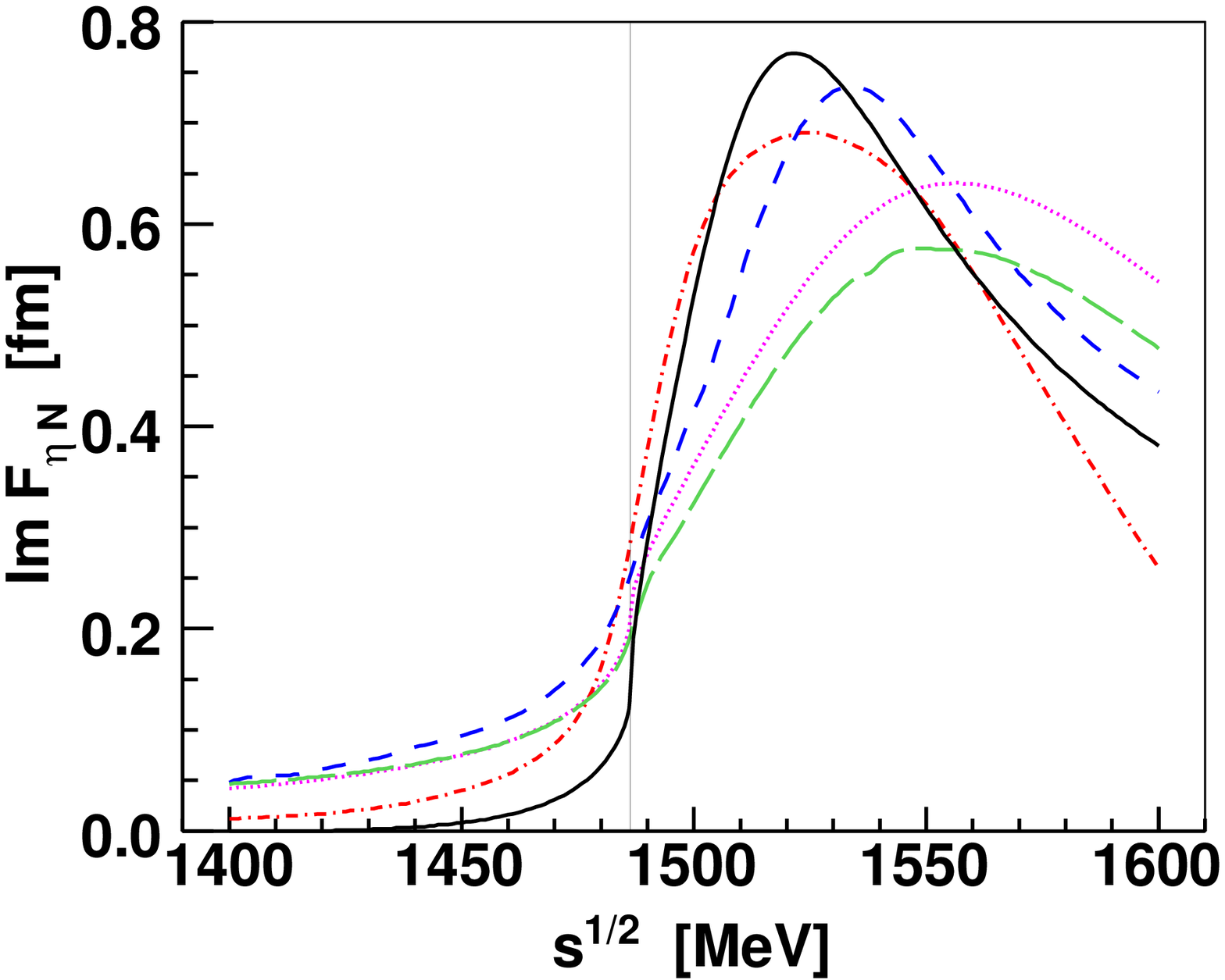}
}
\caption{Real (left panel) and imaginary (right panel) parts of the $\eta N$ 
CMS scattering amplitude $F_{\eta N}(\sqrt{s})$ as a function of the total CMS 
energy $\sqrt{s}$ from five meson-baryon coupled-channel interaction models, 
in decreasing order of Re$\;a_{\eta N}$. Dot-dashed curves: GW \cite{Green:2004tj}; 
solid: CS \cite{Cieply:2013sya}; dotted: KSW \cite{Kaiser:1995cy}; long-dashed: M2 \cite{Mai:2012wy}; 
short-dashed: IOV \cite{Inoue:2001ip}. The thin vertical line denotes the $\eta N$ 
threshold.
}
\label{fig:EtaNampl}       
\end{figure}

In Fig.~\ref{fig:EtaNampl} taken from \cite{Cieply:2013sga} we compare the elastic $\eta N$ amplitude 
generated by our CS model \cite{Cieply:2013sya} with predictions by several other authors
\cite{Kaiser:1995cy, Inoue:2001ip, Mai:2012wy, Green:2004tj}. 
It should be noted that unlike in the $S=0$ sector the various 
theoretical approaches compared here differ conceptually (e.g.~the GW model \cite{Green:2004tj} 
is based on the K-matrix method) as well as in the selection of the fitted experimental data. 
As there is no $\eta$-hydrogen atom either the lack of any direct data at the $\eta N$ threshold 
energy is reflected by varied predictions for the $\eta N$ scattering length $a_{\eta N}$. 
For the real part of the later the considered models provide us with values in a quite a broad 
range from $\Re a_{\eta N} = 0.26$ fm up to $0.96$ fm, while the imaginary part turns out 
to be almost model independent, $\Im a_{\eta N} \approx 0.20 - 0.26$ fm. It appears that a sufficiently 
large value of $\Re a_{\eta N} \gsim 0.7$ fm is a prerequisite for forming quasi-bound $\eta$-nuclear 
states as light as $\eta ^{4}$He \cite{Barnea:2017oyk}. Our comparison presented in Fig.~\ref{fig:EtaNampl} 
demonstrates that the current theoretical predictions are not conclusive in this respect with only 
the models GW \cite{Green:2004tj} and CS \cite{Cieply:2013sya} meeting the criterion. 

%%%%%%%%%%%%%%%%%%%%%%%%%%%%%%%%%%%%%%%%%%%%%%%%%%%%%%%
%%%%%%%%%%%%%%%%%%%%%%%%%%%%%%%%%%%%%%%%%%%%%%%%%%%%%%%

%
%%%%%%%%%%%%%%%%%%%%%%%%%%%%%%%%%%%%%%%%%%%%%%%%%%%%%%%%%%%%%%%%%%%%%%%%%%%%%%%%%%%%%%%%%%%%%%%%%%%%%
\section{Dynamically generated poles}
\label{sec:poles}
%%%%%%%%%%%%%%%%%%%%%%%%%%%%%%%%%%%%%%%%%%%%%%%%%%%%%%%%%%%%%%%%%%%%%%%%%%%%%%%%%%%%%%%%%%%%%%%%%%%%%
%

The chirally motivated approaches predict two poles assigned to the $\Lambda(1405)$ resonance 
\cite{Oller:2000fj}, \cite{Jido:2003cb}. The usual assumption is that one of the poles is formed from 
a resonance in the $\pi\Sigma$ and the other from a quasi-bound state in the $\bar{K}N$ channel 
\cite{Hyodo:2007jq}. In general, the transition amplitudes matrix $F_{ij}$ (with indexes running 
over the channel space) has poles for complex energies $z$ (equal to the meson-baryon CMS energy 
$\sqrt{s}$ on the real axis) if a determinant of the inverse matrix is equal to zero, 
\beq
{\rm det}|F^{-1}(z)| = {\rm det}|V^{-1}(z) - G(z)| = 0  \;\;\; ,
\eeq{eq:det}
where $V$ stands for the potential matrix $V_{ij}$, that is proportional to a coupling matrix $C_{ij}$ 
defined by the underlying chiral symmetry, and the intermediate state Green functions 
are represented by the diagonal $G$ matrix. In the hypothetical zero coupling limit (ZCL), 
in which the non-diagonal inter-channel couplings are switched off ($V_{ij} = 0$ for $i \neq j$), 
the poles persist in the decoupled channels and the condition for a pole of the amplitude becomes
\beq
\prod_n [1/V_{nn}(z) - G_{n}(z)] = 0  \;\;\; .
\eeq{eq:ZCL}
There will be a pole in channel $n$ at a Riemann sheet (RS) [$+/-$] (physical/unphysical) 
if the pertinent $n$-th factor of the product on the r.h.s.~of Eq.~(\ref{eq:ZCL}) 
equals zero. Then, such ZCL poles can be connected with those found in the physical limit 
by following their movement upon increasing gradually the inter-channel couplings 
up to their physical values.

%
%%%%%%%%%%%%%%%%%%%%%%%%%%%%%%%%%%%%%%%%%%%%%%%%%%%%%%%%%%%%%%%%%%%%%%%%%%%%%%%%%%%%%%%%%%%%%%%%%%%%%
\subsection{The strangeness $S=-1$ sector}
\label{sec:KNpoles}
%%%%%%%%%%%%%%%%%%%%%%%%%%%%%%%%%%%%%%%%%%%%%%%%%%%%%%%%%%%%%%%%%%%%%%%%%%%%%%%%%%%%%%%%%%%%%%%%%%%%%
%
The concept of relating the poles origins to their existence in the ZCL was used in \cite{Cieply:2016jby} 
to analyze and compare the pole contents of the current theoretical approaches to the $\bar{K}N$ 
interactions that were reviewed in Section \ref{sec:KbarN}. It follows from Eq.~(\ref{eq:ZCL}) 
that only states with nonzero diagonal couplings $C_{nn} \neq 0$ can generate poles in the ZCL. 
If one considers just the leading order Weinberg-Tomozawa interaction only the $\pi\Sigma$, 
$\bar{K}N$ and $K\Xi$ channels comply with this condition for both, the isoscalar ($I=0$) 
and isovector ($I=1$) states. Though, as some of the models, in particular the Bonn and Murcia ones,  
have sizable NLO contributions, the ZCL poles may not be restricted to just these three channels. 
This can be seen in Table \ref{tab:ZCL} that reviews the findings of \cite{Cieply:2016jby}. 
There, we show, in both isospin sectors, the channels in which a pole assigned to a given resonance 
persists when the ZCL is reached. The two poles assigned to the $\Lambda(1405)$ are shown with 
the indexes $1$ and $2$. In the complex energy plane the $\Lambda_{1}(1405)$ pole is usually 
found at lower energy and further from the real axis than the $\Lambda_{2}(1405)$ pole. For all models 
the origin of the $\Lambda_{1}(1405)$ pole can be traced to a ZCL resonance in the $\pi\Sigma$ channel. 
The $\Lambda_{2}(1405)$ pole couples most strongly to the $\bar{K}N$ channel, so it came as a surprise 
that the pole origin is found in the $\eta\Lambda$ channel for the \MMii{} and \GOii{} models. 
Most likely, the large NLO couplings occurring in these models are responsible for the appearance 
of the $\eta\Lambda$ bound state in the ZCL. The other models that have the ZCL pole in the $\bar{K}N$ 
channel should be preferred if one anticipates a simplified picture of a $\bar{K}N$ bound state 
submerged in the $\pi\Sigma$ continuum \cite{Hyodo:2007jq}. 

%%%%%%%%%%%%%%%%%%%%%%%%%%%%%%%%%%%%%%%%%%%%%%%%%%%%%%%
\begin{table}[htb]
\centering
\caption{The origins (channels) of the poles generated by the models considered in the $\bar{K}N$ sector 
for which the poles are found when inter-channel couplings are switched off.}
\begin{tabular}{cc|cccccc}
&\multicolumn{7}{c}{\hspace{2cm}Models} \\
&                     &   \CSnlo{}   &   \IHWnlo{}  &   \GOi{}      &  \GOii{}      &  \MMii{}      &  \MMiv{}    \\
\cline{2-8}
\multirow{5}{*}{\rotatebox[origin=c]{90}{Resonances}}&$\Lambda_{1}(1405)$  &  $\pi\Sigma$ &  $\pi\Sigma$ &  $\pi\Sigma$  & $\pi\Sigma$   & $\pi\Sigma$   & $\pi\Sigma$ \\
&$\Lambda_{2}(1405)$  &  $\bar{K}N$  &  $\bar{K}N$  &  $\bar{K}N$   & $\eta\Lambda$ & $\eta\Lambda$ & $\bar{K}N$  \\
&$\Lambda(1670)$      &  $K\Xi$      &     ---      &  $K\Xi$       & $K\Xi$        &     ---       & $K\Xi$      \\
&$\bar{K}N(I=1)$      &  $\bar{K}N$  & $\eta\Sigma$ &  $\bar{K}N$   & $\bar{K}N$    &     ---       &   ---       \\
&$\Sigma(1750)$       &  $K\Xi$      &     ---      &  $K\Xi$       & $K\Xi$        &     ---       & $K\Xi$      
\end{tabular}
\label{tab:ZCL}       
\end{table}
%%%%%%%%%%%%%%%%%%%%%%%%%%%%%%%%%%%%%%%%%%%%%%%%%%%%%%%

In the isoscalar sector the models can also account for the $\Lambda(1670)$ resonance 
that emerges from the $K\Xi$ pole found in the ZCL. It was argued in Ref.~\cite{Cieply:2016jby} 
that an appearance of such pole is related to a particular condition imposed on a subtraction constant 
(or an inverse range in case of the Prague approach). If the condition is not met, the pole is missing 
as happens for the \IHWnlo{} and \MMii{} models. One should note, however, that with an exception  
of the Murcia approach the other approaches did not aim at describing the experimental data 
in the $\Lambda(1670)$ energy region, so it is not surprising that the pole is either completely 
missing or not at an appropriate position in those models.

Similarly, in the isovector sector the models can provide a pole which can be related 
to the $\Sigma(1750)$ resonance and the origin of this pole can be traced to the $K\Xi$ virtual (or bound) 
state in the ZCL. Several of the discussed models also predict an isovector $\bar{K}N$ 
pole located below the $\bar{K}N$ threshold at the Riemann sheet which is physical in the $\pi\Sigma$ 
and unphysical in the $\bar{K}N$ channel (it would be the third Riemann sheet if only these 
two channels were coupled). This pole emerges from an isovector $\bar{K}N$ virtual 
state generated in the ZCL by the Prague and Murcia models, though the Kyoto-Munich model has 
it in the $\eta\Sigma$ channel. We note that an existence of this pole was already witnessed 
in Refs.~\cite{Oller:2000fj}, \cite{Jido:2003cb} and \cite{Cieply:2011fy}. It is understood 
that it relates to the cusp structure in the energy dependence of the elastic $K^{-}n$ amplitude 
obtained for both, the \CSnlo{} and the \IHWnlo{} models as seen in Fig.~\ref{fig:KNampl}.

%
%%%%%%%%%%%%%%%%%%%%%%%%%%%%%%%%%%%%%%%%%%%%%%%%%%%%%%%%%%%%%%%%%%%%%%%%%%%%%%%%%%%%%%%%%%%%%%%%%%%%%
\subsection{The strangeness $S=0$ sector}
\label{sec:etaNpoles}
\begin{figure}[htb]    
\centering
\includegraphics[width=0.6\textwidth]{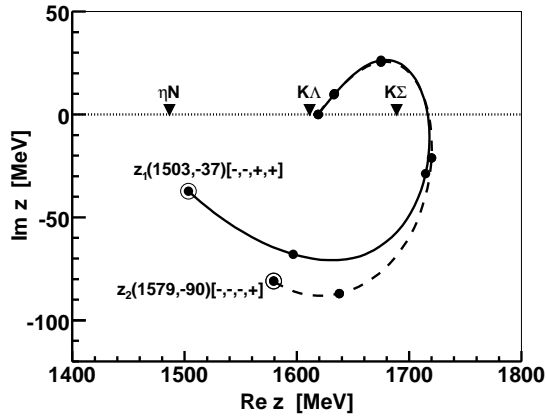}
\caption{Movement of the poles $z_1$ and $z_2$ upon gradually switching off the inter-channel couplings. 
The positions of the poles in a physical limit are encircled and marked by the labels that also denote 
the Riemann sheets the poles are located on. The small dots mark positions of the poles for the scaling 
factors from $x=0$ (zero coupling limit) to $x=1$ (physical limit) in steps of $0.2$.}
\label{fig:ZCL}
\end{figure}

%%%%%%%%%%%%%%%%%%%%%%%%%%%%%%%%%%%%%%%%%%%%%%%%%%%%%%%%%%%%%%%%%%%%%%%%%%%%%%%%%%%%%%%%%%%%%%%%%%%%%
%
A similar comparative analysis of the poles origins in the $\eta N$ related $S=0$ sector is missing. 
Among the models discussed in Section \ref{sec:etaN} just the authors of the CS model \cite{Cieply:2013sya} 
studied the pole movements when going into the ZCL. For strangeness $S=0$ and isospin $I=1/2$  
only the $\pi N$ and $K\Sigma$ channels have nonzero diagonal couplings. Since the $\pi N$ 
threshold is too far below the $\eta N$ one the only feasible option (barring large NLO contributions) 
to generate dynamically a pole at energies above 1500 MeV appears to be the $K\Sigma$ channel 
with a diagonal coupling strong enough to generate a virtual state. This is confirmed in 
Fig.~\ref{fig:ZCL} taken from \cite{Cieply:2013sya} we show here to demonstrate the movement 
of the poles on two different Riemann sheets labeled by signs of the imaginary parts 
of meson-nucleon CMS momenta in all involved channels. In this notation [+,+,+,+] represents 
the physical RS and the signs are in order of the channel thresholds. Both pole trajectories 
shown in Fig.~\ref{fig:ZCL} emerge from the same $K\Sigma$ virtual state in the ZCL. The trajectory 
developing on the [+,+,-,-] RS (continuous line, $z_1$ pole) reaches the position assigned 
to the $N^{\star}(1535)$ resonance in the physical limit. The shadow pole that develops 
on the [+,+,+,-] RS (dashed line, $z_2$ pole) can be related to the $N^{\star}(1650)$ resonance. 
Although, the $z_2$ pole energy in the physical limit is significantly off the experimental one, 
the position of the pole can be tuned by varying the model parameters. 

Concerning the $N^{\star}(1535)$ resonance we would also like to stress that it is far from clear 
that it should be generated dynamically. The fact that the dynamically generated $N^{\star}(1535)$ 
pole emerges from the $K\Sigma$ virtual state located at much higher energies indicates 
that the parameters of the CS model might be stretched beyond their feasible limits. 
However, the coupled channels model of Ref.~\cite{Cieply:2013sya} has no other choice 
to accommodate the resonant shape observed in the $\eta N$ formation cross 
section. Thus, one could argue in favor of considering a genuine excited quark state 
to represent the resonance instead of generating it dynamically. 
We also note that a satisfactory description of both the $N^{\star}(1535)$ and $N^{\star}(1650)$ 
resonances can be achieved by considering the resonant components of the $\pi\pi N$ system 
in addition to the pseudoscalar-baryon coupled channels discussed here, see e.g.~Refs.~\cite{Kamano:2013iva}
and \cite{Garzon:2014ida}.

%
%%%%%%%%%%%%%%%%%%%%%%%%%%%%%%%%%%%%%%%%%%%%%%%%%%%%%%%%%%%%%%%%%%%%%%%%%%%%%%%%%%%%%%%%%%%%%%%%%%%%%
\section{Summary}
\label{sec-summary}
%%%%%%%%%%%%%%%%%%%%%%%%%%%%%%%%%%%%%%%%%%%%%%%%%%%%%%%%%%%%%%%%%%%%%%%%%%%%%%%%%%%%%%%%%%%%%%%%%%%%%
%

We have demonstrated that the current theoretical models provide quite varied predictions for 
the elastic $\bar{K}N$ and $\eta N$ amplitudes at subthreshold energies and in case of the $\eta N$ 
system at the threshold too. This observation is intriguing since the same models describe about equally 
well the available experimental data including the shape of the near threshold $\Lambda(1405)$ and 
$N^{\star}(1535)$ resonances observed in the $\pi \Sigma$ and $\eta N$ spectra, respectively. 
Additional experimental constrains on the discussed theoretical models can be provided by studying 
the impact of nuclear matter on the coupled-channels meson-baryon dynamics. In particular, it was 
already revelaed in \cite{Friedman:2016rfd} that only the $\bar{K}N$ amplitudes generated by the 
\IHWnlo{} and \CSnlo{} models are consistent with the systematics of kaonic atom data 
and reproduce the observed fraction of antikaon absorption on single nucleons. It is worth noting 
that the Murcia and Bonn approaches that do not pass this additional in-medium test are those 
characterized by large NLO contributions, a feature that distinguishes them from 
the Kyoto-Munich and Prague ones.

The origin of resonances generated dynamically in meson-baryon coupled channels systems can be traced 
to poles emerging due to sufficiently strong interactions in the decoupled channels with inter-channel 
couplings switched off. These ZCL poles are then moved into their positions found in the physical limit 
when the inter-channel couplings are gradually turned on. This concept provides us with additional 
insights on the conditions under which dynamical resonances are formed. The presented contribution 
reviews our recent findings for the poles assigned to the $\Lambda^{\star}$ and $\Sigma^{\star}$ 
resonances observed in the meson-baryon $S=-1$ channels coupled to the $\bar{K}N$ system,  
and for the poles assigned to the $N^{\star}$ resonances in the $S=0$ channels related 
to the $\eta N$ system. Besides confirming and analyzing the double pole structure 
of the $\Lambda(1405)$ we also note that some chirally motivated models predict an existence 
of the isovector pole at energies not far below the $\bar{K}N$ threshold. In the $S=0$ sector,
the chirally motivated coupled-channels dynamics can account for the $N^{\star}(1535)$ 
and $N^{\star}(1650)$ resonances, though the dynamical status of the first one might be questioned.

\section*{Acknowledgements}
This work was supported by the GACR Grant No.~P203/15/04301S.
The author also acknowledges a collaboration with M.~Mai, U.-G.~Mei{\ss}ner and J.~Smejkal who contributed 
to Refs.~\cite{Cieply:2016jby} and \cite{Cieply:2013sya} the current presentation is based on.

%
% BibTeX or Biber users please use (the style is already called in the class, ensure that the "woc.bst" 
% style is in your local directory)
% \bibliography{name or your bibliography database}
%
% Non-BibTeX users please use
%

\end{document}